\documentclass[conference]{IEEEtran}
\IEEEoverridecommandlockouts
\usepackage{cite}
\usepackage{amsmath,amssymb,amsfonts}
\usepackage{algorithmic}
\usepackage{graphicx}
\usepackage{textcomp}
\usepackage{xcolor}
\usepackage{hyperref}
\def\BibTeX{{\rm B\kern-.05em{\sc i\kern-.025em b}\kern-.08em
    T\kern-.1667em\lower.7ex\hbox{E}\kern-.125emX}}

\begin{document}

\title{ Angular upsampling in diffusion MRI using contextual \emph{HemiHex} sub-sampling in \emph{q}-space\\
}

\author{\IEEEauthorblockN{Abrar Faiyaz}
\IEEEauthorblockA{\textit{Department of ECE } \\
\textit{University of Rochester}\\
Rochester, US \\
afaiyaz@ur.rochester.edu}
\and
\IEEEauthorblockN{Md Nasir Uddin}
\IEEEauthorblockA{\textit{Department of Neurology} \\
\textit{University of Rochester}\\
Rochester, US \\
Nasir\_Uddin@URMC.rochester.edu}
\and
\IEEEauthorblockN{Giovanni Schifitto}
\IEEEauthorblockA{\textit{Department of Neurology} \\
\textit{University of Rochester}\\
Rochester, US \\
Giovanni\_Schifitto@URMC.Rochester.edu}

}

\maketitle

\begin{abstract}
Artificial Intelligence (Deep Learning(DL)/ Machine Learning(ML)) techniques are widely being used to address and overcome all kinds of ill-posed problems in medical imaging which was or in fact is seemingly impossible. Reducing gradient directions but harnessing high angular resolution(HAR) diffusion data in MR that retains clinical features is an important and challenging problem in the field. While the DL/ML approaches are promising, it is important to incorporate relevant context for the data to ensure that maximum prior information is provided for the AI model to infer the posterior. In this paper, we introduce \emph{HemiHex}(HH) subsampling to suggestively address training data sampling on q-space geometry, followed by a nearest neighbor regression training on the HH-samples to finally upsample the dMRI data. Earlier studies has tried to use regression for up-sampling dMRI data but yields performance issues as it fails to provide structured geometrical measures for inference.

Our proposed approach is a geometrically optimized regression technique which infers the unknown q-space thus addressing the limitations in the earlier studies. 
\end{abstract}

\begin{IEEEkeywords}
HemiHex, angular, upsampling, diffusion, MRI, deep learning, machine learning, Subsampling, q-space, delaunay, triangulation, low angular resolution, lar, har
\end{IEEEkeywords}

\section{Introduction}
Harnessing high angular resolution(HAR) with reduced number of gradients in diffusion
data in MR that retains clinical features is an important and challenging problem in the field\cite{chen_angular_2018}. Interpolation techniques to estimate DWI signals for Fiber oreientation density(fod) estimation and tractography with dMRI has shown promising result in simulation and \emph{in-vivo}\cite{alaya_quantitative_2019,ben_alaya_fast_2017,ben_alaya_comparison_2022}. We hypothesized that if the upsampling procedure can be guided through a Delanuy Triangulation sampling scheme, the nearest neighbor geometry could potentially be generalized in terms of diffusion signals. 

The dataset provided in the MICCAI Challenge held in the CDMRI workshop 2022 (QuaD22) has geometrically optimized sampling scheme for both upsampled and low angular resolution q-space data and this enables us to generalize subsamples of the q-space and finally estimate unknowns in the geometry.

\section{Method}

\subsection{Optimized sampling in q-space}
We have utilized Iterative Maximum
Overlap Construction (IMOC) and 1 Opt greedy method for optimizing gradient directions.\cite{cheng_designing_2014, cheng_single-_2018} The gradient nodes in LAR were considered in such a way that they have the best angular coverage. Then the HAR gradients can be optimized considering LAR gradients to be a subset of the HAR protocol. The protocol optimization was done using dMRItool in MATLAB.

\subsection{HemiHex Subsampling in q-space}
\begin{figure}
\centering
\includegraphics[height=2in]{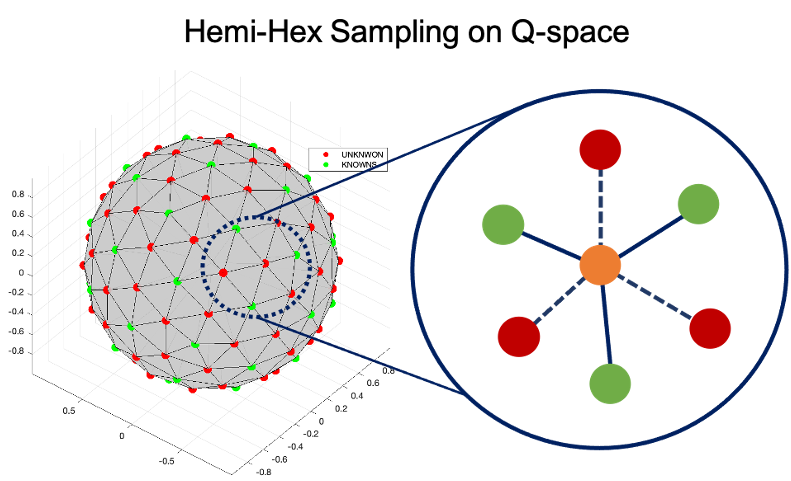}
\caption{Hemi-hex subsampling on q-space, the center node is an unknown surrounded with 3 known and unknowns.}
\label{hemihex}
\end{figure}

HemiHex (HH) Subsampling can be regarded as subsampling the q-space centering on an unknown node in such a way that the known and unknown q-space nodes fall on approximated hexagon nodes, alternating knowns and unknowns.
Hemi-Hex interpolation refers to the regression of unknown center node on a hemi-hex sample.
The subsampling scheme results in more training data points per subject and thus requires less subjects to be trained contextually. (Figure-\ref{hemihex})

\subsection{HemiHex Subsampling Prerequisites }
\begin{itemize}
\item Both the LAR and HAR images must have optimized sampled schemes in q space for diffusion protocol. Optimization makes sure that the gradients are equidistant and the hexagonal formation is possible between the known and unknown samples.

\item The LAR gradients used must be common to HAR gradients and should be a subset of HAR grads. Number of gradients in LAR must be 3 times the number of gradients in the HAR. 
\end{itemize}

\begin{figure*}
\centering
\includegraphics[height=4in]{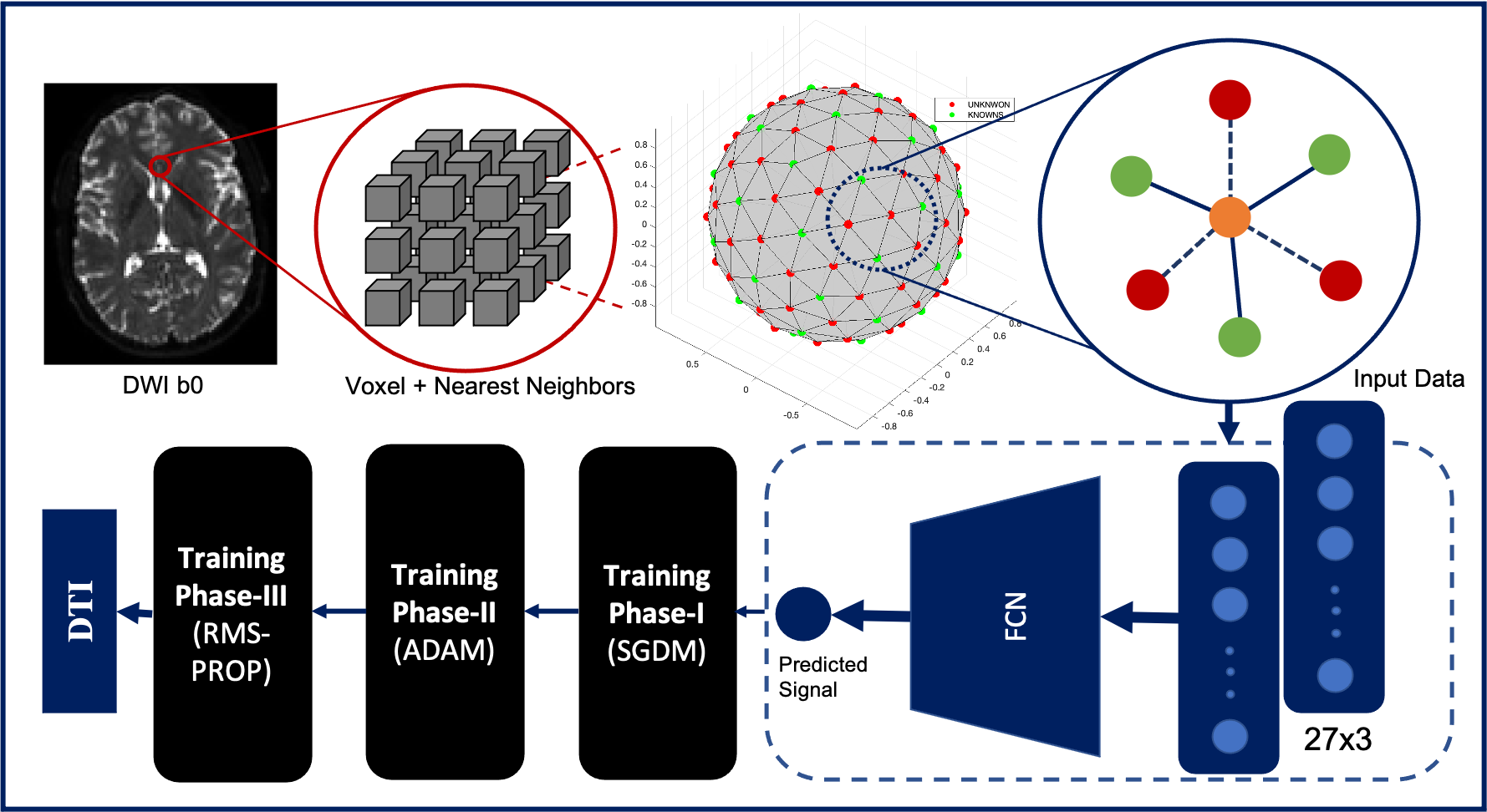}
\caption{Schematics of angular upsampling in diffusion MR using contextual HemiHex sub-sampling in q-space}\label{gabs}
\end{figure*}
\subsection{Mahcine Learning Model }
A fully connected network (FCN) is used with 27x3 input layer and one node for diffusion weighted signal output for each gradient. The schematic diagram of the approach is shown in Figure-\ref{gabs}. HemiHex subsampling generates training and validation data required for training the FCN. Spatial neighbors are incorporated in the input domain of the learner.

\begin{figure}
\centering
\includegraphics[height=2in]{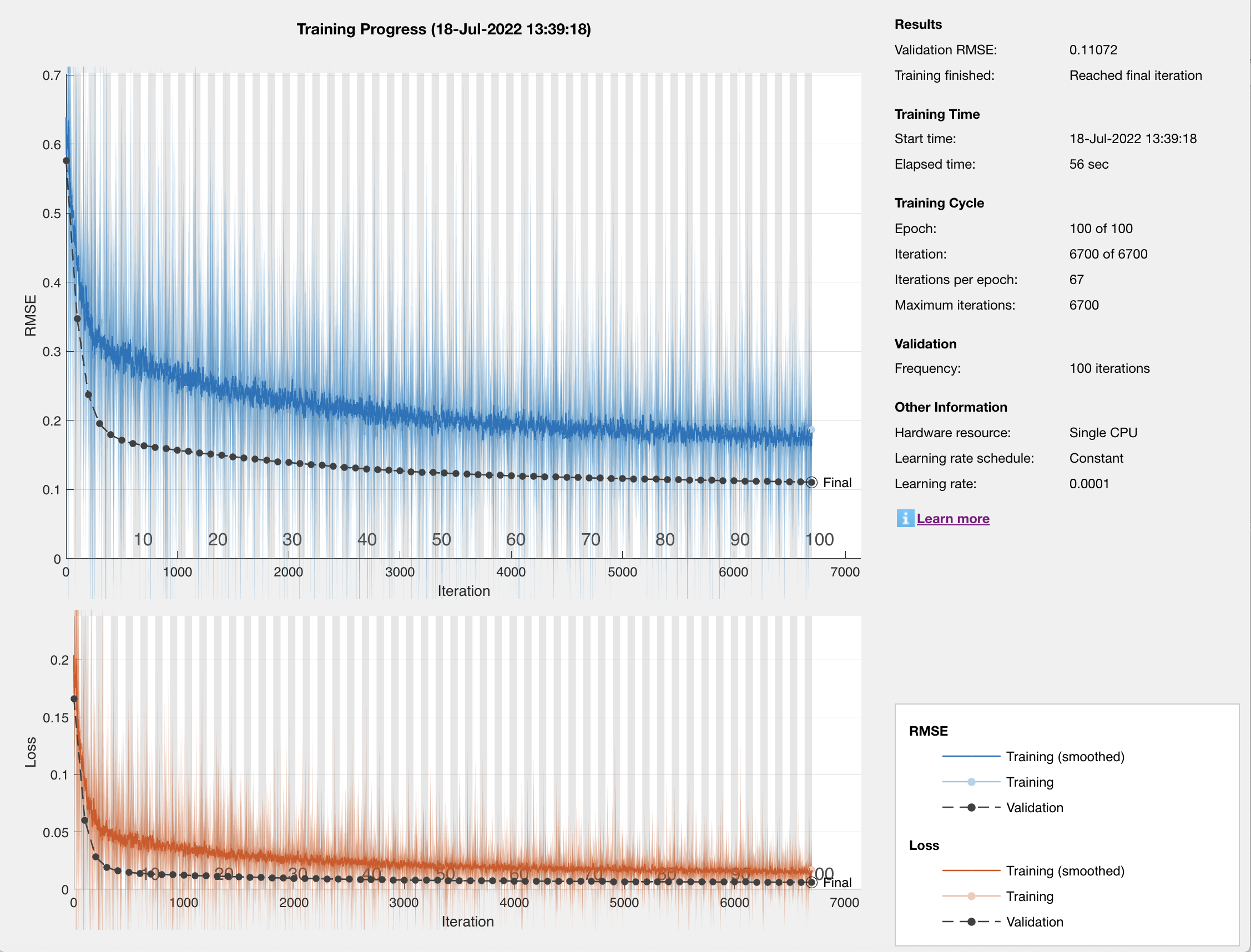}
\caption{Training Phase-1 with stochastic gradient descent with momentum details highlights initial loss and RMSE for training and validation iterations.}
\label{sgdm}
\end{figure}

\begin{figure}
\centering
\includegraphics[height=2in]{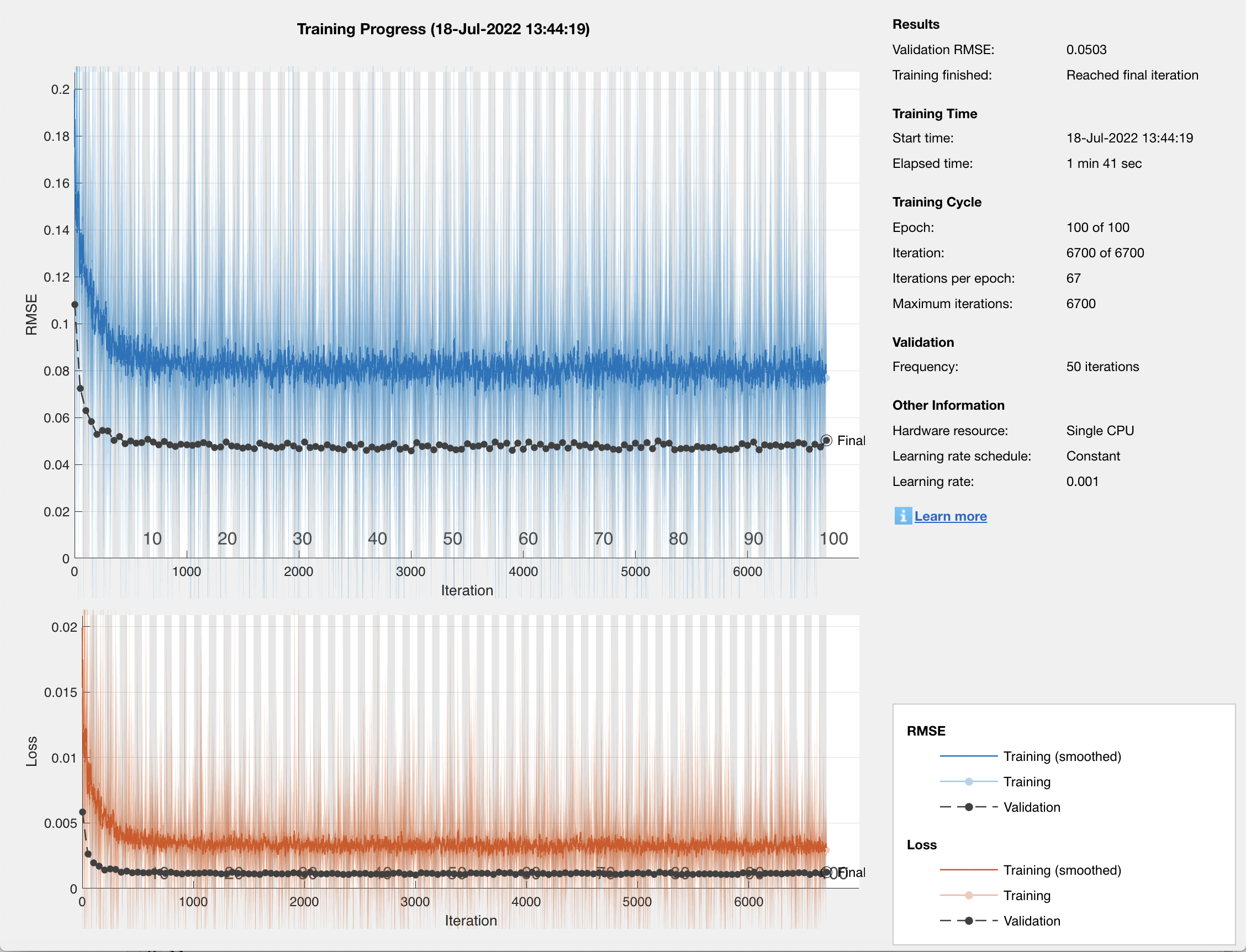}
\caption{Training Phase-2 with ADAM highlights secondary loss and RMSE for training and validation iterations.}
\label{adam}
\end{figure}

\begin{figure}
\centering
\includegraphics[height=2in]{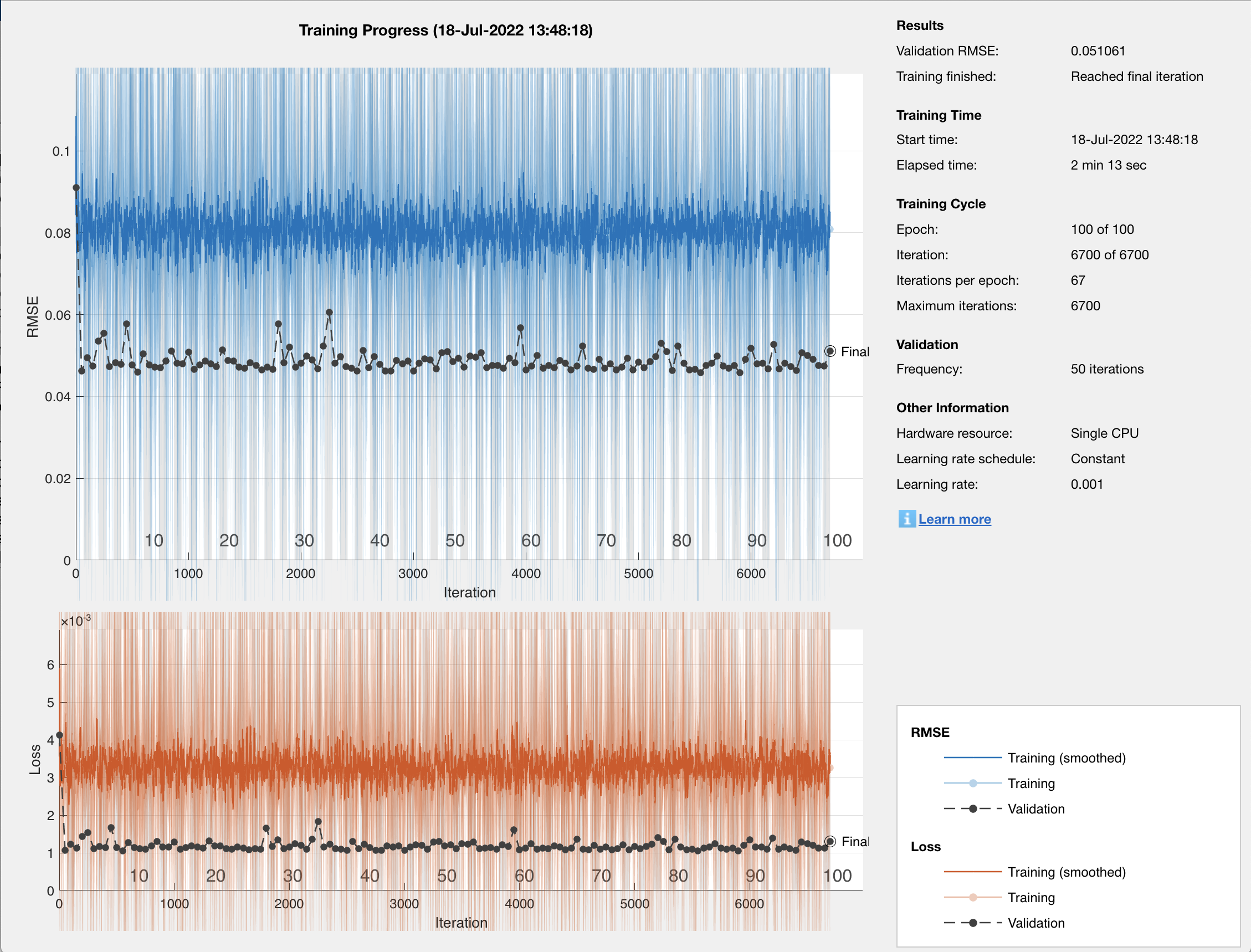}
\caption{Training Phase-3 with rmsprop highlights final loss and RMSE for training and validation iterations.}\label{rmsprop}
\end{figure}

The network is trained in three different phases with stochastic gradient descent with momentum, adam and rmsprop algorithms respectively. Successive iteration and details are reported on Figure-\ref{sgdm},\ref{adam},\ref{rmsprop}.

ROIs were selected from GM, WM and merging areas from three subjects to obtain the training data. Validation data was similarly extracted from five other subjects. In total, eight subjects are used for training and validating the network for DWI signal prediction. Since the training takes place in regressing the fourth dimension (the dwi signals), sufficient training data is sampled from eight healthy subjects.

The input samples are 27 dimensional as each centering voxel posses 26 neighbours. Based on Delaneuy triangulation geometry, for each unknown signal we always have 3 knowns on the sphere. The architecture is expected to be rotation invariant because the training being generalized on the sphere due to the inherent sampling geometry of the provided data. For each voxel, diffusion gradient signals for 40 unknown gradients are estimated.

\subsection{Estimation Philosophy} The network estimates DW signals for the unknown directions trained on the variable yet close geometric patterns. The final metric of evaluation are the DTI metrics. Once the data is upsampled, DTI is applied on the data and the final result is generated. The results are not shown in this article.
\subsection{Dataset}
Two datasets in the study were obtained through MICCAI Challenge 2022 - \href{https://www.lpi.tel.uva.es/quad22/index.html}{QuaD22 Website}

\begin{itemize}
    \item \textbf{Training dataset:}  Diffusion-weighted MRI dataset acquired with 61 gradient directions at b=1000 s/mm\textsuperscript{2} coming from 60 healthy controls. The sampling scheme allows the 61 gradient directions to be easily subsampled to 21 gradients. (Resolution- 1.875x1.875x2 mm\textsuperscript{3})
    \item \textbf{Migraine dataset:} A set of 50 Chronic migraine (CM) and 50 episodic migraine (EM) patients, all acquired in a subsampled scenario with 21 gradient directions and b=1000 s/mm\textsuperscript{2}.(Resolution- 1.875x1.875x2 mm\textsuperscript{3})
\end{itemize}

\emph{Note: Although 60 healthy controls were provided from the challenge website, only 3 was used for training purposes. Others were used for validation and testing.}
\section{Optimization}
For training the nearest neighbor regression network, we have used three different optimization algorithms. Stochastic gradient descent with momentum (sgdm), adam and rmsprop algorithms successively were applied on independent training data points to minimize the loss function. Three phase training had the same objective function for minimization, 
$$ MSE=\sum_{i=1}^{D}(x_i-y_i)^2 $$

where, $ D $ is the number of total gradient directions for a voxel in x-space. $ x_i $ denotes the original gradient signal and $ y_i $ is the predicted gradient signal inferred from the subsampled input.
\begin{figure}
\centering
\includegraphics[height=2in]{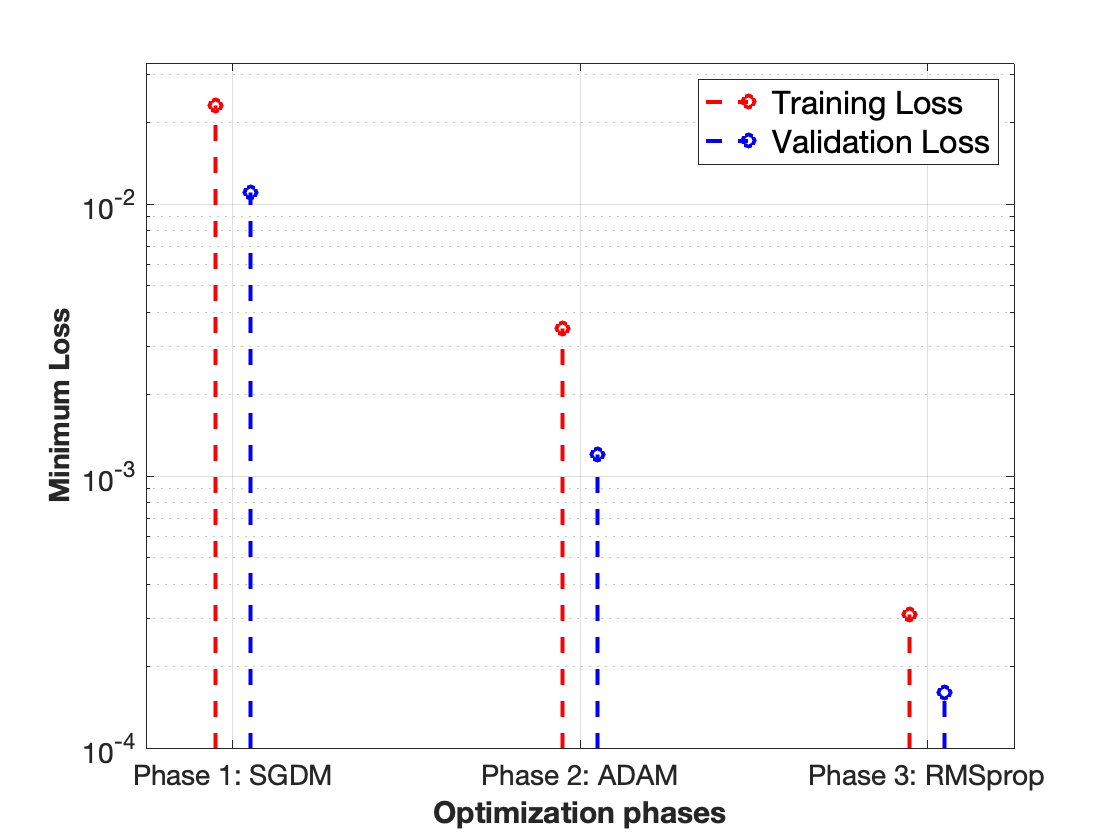}
\caption{Three step optimization reduces the error in logscale.}\label{optmize}
\end{figure}

\section{Discussion}
Angular Upsampling  is an ill-posed inverse problem which can be made stable with the help of tessellated geometries designed in this work through sampling.  Tessellated geometries ensure rotation invariant learning on the sphere (q-space). And the neighboring voxels provide local spatial prior to learn the tissue property for a specific gradient. This makes sure high resolution  counterpart can retain clinically relevant information through upsampling.

The implemented approach requires significantly less computational resources than general CNN based architectures. (We have used core i7 CPU with 16GB memory used for training). Which took less than 10 minutes for training.

The approach is highly efficient with training resources which yields important advantage for clinical applications and research. As less as 2/3 subjects is enough for training and acquire the contextual information for the gradients.

\section{Conclusion}
We proposed a fully connected regression network integrating the philosophy of geometric distribution and interpolation of DWI data. The low resource need and ease of computation makes the approach more suitable to be used in dMRI clinical studies.

\section*{Acknowledgements \& Note}

The work is a participant of MICCAI Challenge 2022, organized by CDMRI workshop, QuaD22. The provided dataset result or the competition result is not released or shown in this work in agreement with the organizers. Only the developed method is presented in the article.






\bibliographystyle{IEEEtran}
\bibliography{conference_101719}  %



\end{document}